\documentclass[runningheads]{llncs}
\usepackage{cvprabbreviation}
\usepackage{graphicx}
\usepackage{comment}
\usepackage{amsmath,amssymb} 
\usepackage{color}
\usepackage[figuresright]{rotating}
\usepackage{pdflscape}
\usepackage{adjustbox}

\usepackage{caption}
\usepackage{subfigure}
\usepackage{lipsum}
\usepackage{epstopdf}
\usepackage{array}
\usepackage{multirow}
\usepackage{cite}
\usepackage[american]{babel}
\usepackage{bbding} 
\usepackage[breaklinks=true,letterpaper=true,colorlinks,urlcolor=black,bookmarks=false]{hyperref}

\newcommand{\widthscalefive}{0.14}

\begin{document}
\pagestyle{headings}
\mainmatter
\def\ECCVSubNumber{xxx}  

\newcommand*{\affaddr}[1]{#1} 
\newcommand*{\affmark}[1][*]{\textsuperscript{#1}}
\renewcommand*{\email}[1]{\texttt{#1}}
\renewcommand*{\institute}[1]{#1}
\title{Deep Adaptive Inference Networks for\\ Single Image Super-Resolution} 


\titlerunning{Deep Adaptive Inference Networks for SISR}
%
\author{%
Ming Liu\affmark[1] \and
Zhilu Zhang\affmark[1] \and
Liya Hou\affmark[1] \and
Wangmeng Zuo\affmark[1 (\Envelope)] \and
Lei Zhang\affmark[2]\\
\institute{\small \affmark[1]Harbin Institute of Technology, Harbin, China}\\
\institute{\small \affmark[2]The Hong Kong Polytechnic University, Hong Kong, China}\\
\email{\small csmliu@outlook.com,}
\email{\small cszlzhang@outlook.com,}
\texttt{\small \href{mailto:h\_liya@outlook.com}{h\_liya@outlook.com},}
\email{\small wmzuo@hit.edu.cn,}
\email{\small cslzhang@comp.polyu.edu.hk}}

\authorrunning{M. Liu, Z. Zhang, L. Hou, W. Zuo and L. Zhang.}

\maketitle

\vspace{-1.5em}
\begin{abstract}
Recent years have witnessed tremendous progress in single image super-resolution (SISR) owing to the deployment of deep convolutional neural networks (CNNs).
For most existing methods, the computational cost of each SISR model is irrelevant to local image content, hardware platform and application scenario.
Nonetheless, content and resource adaptive model is more preferred, and it is encouraging to apply simpler and efficient networks to the easier regions with less details and the scenarios with restricted efficiency constraints.
In this paper, we take a step forward to address this issue by leveraging the adaptive inference networks for deep SISR (AdaDSR).
In particular, our AdaDSR involves an SISR model as backbone and a lightweight adapter module which takes image features and resource constraint as input and predicts a map of local network depth.
Adaptive inference can then be performed with the support of efficient sparse convolution, where only a fraction of the layers in the backbone is performed at a given position according to its predicted depth.
The network learning can be formulated as the joint optimization of reconstruction and network depth losses.
In the inference stage, the average depth can be flexibly tuned to meet a range of efficiency constraints.
Experiments demonstrate the effectiveness and adaptability of our AdaDSR in contrast to its counterparts (\eg, EDSR and RCAN).

\vspace{-0.5em}
\end{abstract}

\vspace{-2em}
\section{Introduction}\label{sec:intro}
\vspace{-0.5em}
Image super-resolution aims at recovering high-resolution (HR) image from its low-resolution (LR) counterpart, is a representative low-level vision task with many real-world applications such as medical imaging~\cite{shi2013cardiac}, surveillance~\cite{zou2011very} and entertainment~\cite{old_film}.
Recently, driven by the development of deep convolutional neural networks (CNNs), tremendous progress has been made in single image super-resolution (SISR).
On the one hand, the quantitative performance of SISR has been continuously improved by many outstanding representative models such as SRCNN~\cite{SRCNN}, VDSR~\cite{VDSR}, SRResNet~\cite{SRResNet}, EDSR~\cite{EDSR}, RCAN~\cite{RCAN}, SAN~\cite{SAN}, \etc.
On the other hand, considerable attention has also been given to handle several other issues in SISR, including visual quality~\cite{SRResNet}, degradation model~\cite{DPSR}, and blind SISR~\cite{zhang2019multiple}.

\begin{figure}[t]
    \centering
    \begin{minipage}{0.23\linewidth}
        \subfigure[LR image]{
        \begin{minipage}{\linewidth}
            \vspace{0.5em}
            \includegraphics[width=\linewidth, height=5.5em]{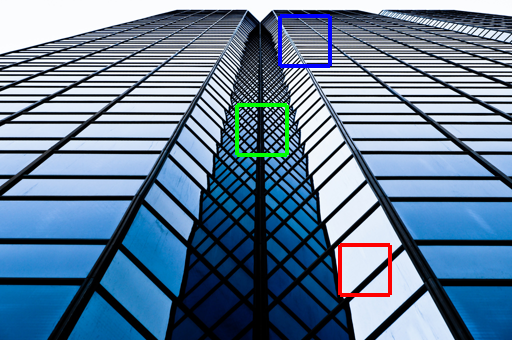}
        \end{minipage}}
        \vspace{-.8em}
        \subfigure[Depth map]{
        \begin{minipage}{\linewidth}
            \includegraphics[width=\linewidth, height=5.5em]{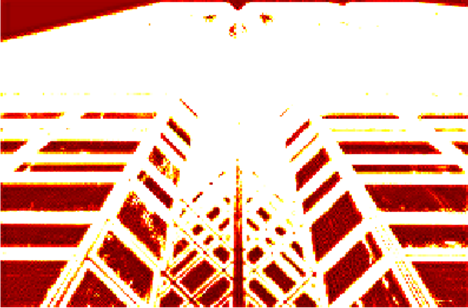}
        \end{minipage}}
    \end{minipage}
    \subfigure[Depth - Performance]{
    \begin{minipage}{0.35\linewidth}
        \includegraphics[width=\linewidth, height=13em]{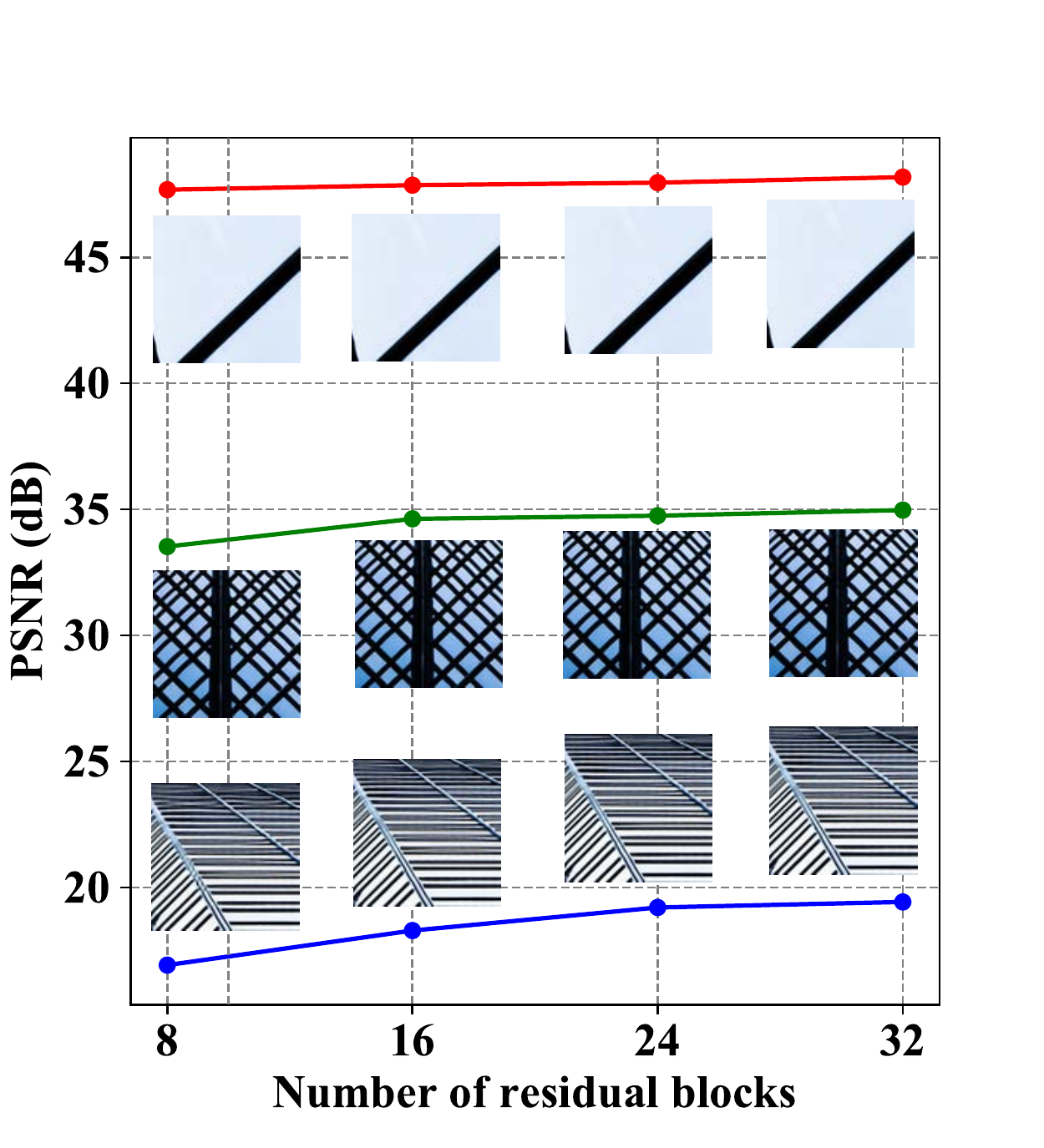}
    \end{minipage}}
    \subfigure[Comparison on Set5 (x2)]{
    \begin{minipage}{0.33\linewidth}
        \includegraphics[width=\linewidth, height=13em]{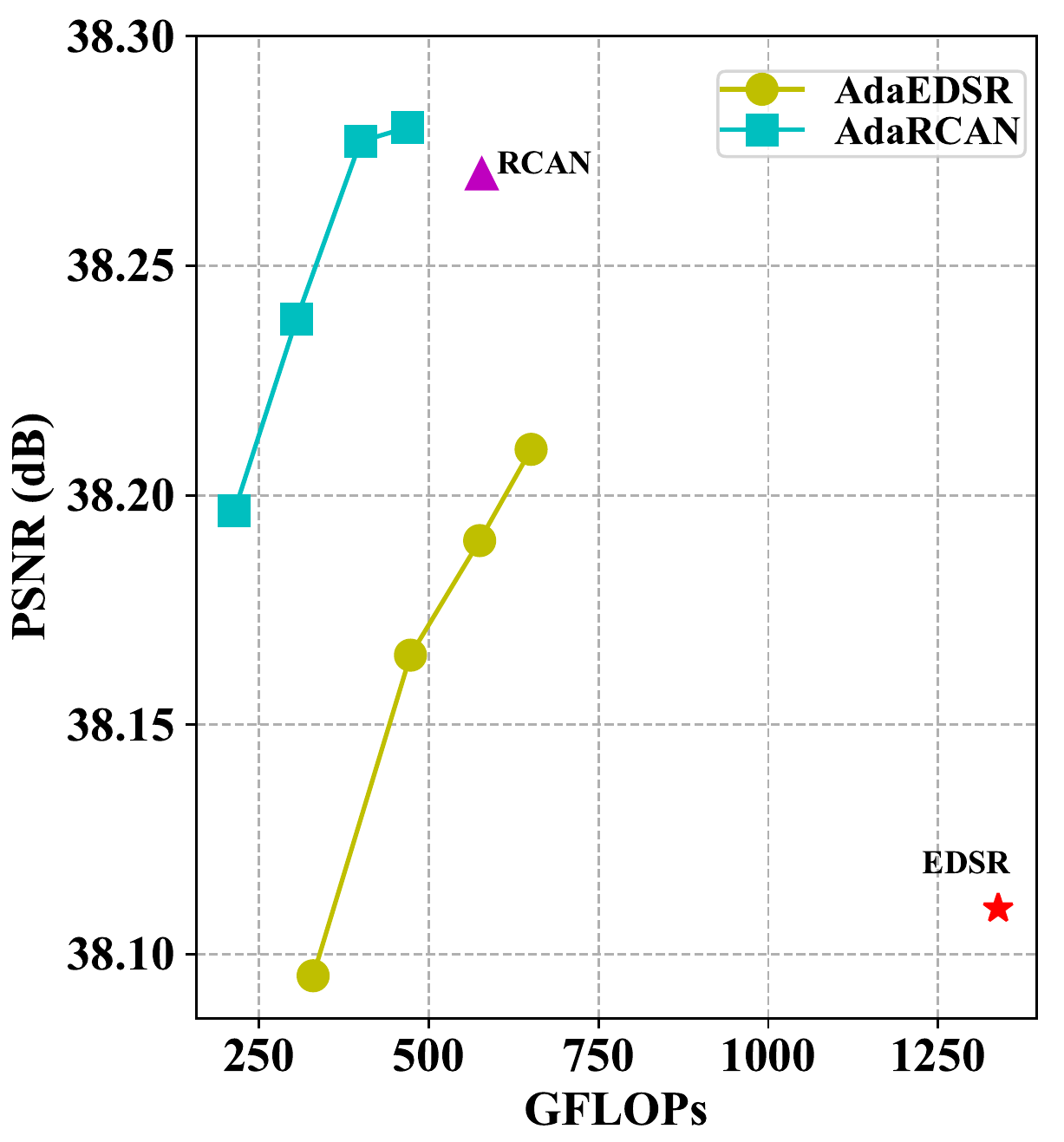}
    \end{minipage}}
    \vspace{-0.5em}
    \caption{\footnotesize Illustration of our motivation and performance. (a) and (b) show an LR image and the depth map predicted by our AdaDSR model, and three representative patches with various SISR difficulty are marked out. (c) explores the performance of EDSR models with different number of residual blocks on these patches. In (d), we compare two versions of our AdaDSR against their backbones on Set5 dataset. Please zoom in for better observation, and refer to the supplementary materials for more comparison on other conditions.}\label{fig:intro}
    \vspace{-2em}
\end{figure}

Albeit their unprecedented success of SISR, for most existing networks, the computational cost of each model is still independent to image content and application scenarios.
Given such an SISR model, once the training is finished, the inference process is deterministic and only depends on the model architecture and the input image size.
Actually, instead of deterministic inference, it is inspiring to make the inference to be adaptive to local image content.
To illustrate this point, Fig.~\ref{fig:intro}(c) shows the SISR results of three image patches using EDSR~\cite{EDSR} with different numbers of residual blocks.
It can be seen that EDSR with 8 residual blocks is sufficient to super-resolve a smooth patch with less textures.
In contrast, at least 24 residual blocks are required for the patch with rich details.
Consequently, treating the whole image equally and processing all regions with identical number of residual blocks will certainly lead to the waste of computation resource.
Thus, it is encouraging to develop the spatially adaptive inference method for better tradeoff between accuracy and efficiency.

Moreover, the SISR model may be deployed to diverse hardware platforms.
Even for a given hardware device, the model can be run under different battery conditions or workloads, and has to meet various efficiency constraints.
One natural solution is to design and train numerous deep SISR models in advance, and dynamically select the appropriate one according to the hardware platform and efficiency constraints.
Nonetheless, both the training and storage of multiple deep SISR models are expensive, greatly limiting their practical applications to the scenarios with highly dynamic efficiency constraints.
Instead, we suggest to address this issue by further making the inference method to be adaptive to efficiency constraints.

To make the learned model to adapt to local image content and efficiency constraints, this paper presents a kind of adaptive inference networks for deep SISR, \ie, AdaDSR.
Considering that stacked residual blocks have been widely adopted in the representative SISR models~\cite{SRResNet, EDSR, RCAN}, the AdaDSR introduces a lightweight adapter module which takes image features as the input and produces a map of local network depth.
Therefore, given a position with the local network depth $d$, only the first $d$ blocks are required to be computed in the testing stage.
Thus, our AdaDSR can apply shallower networks for the smooth regions (\ie, lower depth), and exploit deeper ones for the regions with detailed textures (\ie, higher depth), thereby benefiting the tradeoff between accuracy and efficiency.
Taking all the positions into account, sparse convolution can be adopted to facilitate efficient and adaptive inference.

We further improve AdaDSR to be adaptive to efficiency constraints.
Note that the average of depth map can be used as an indicator of inference efficiency.
For simplicity, the efficiency constraint on hardware platform and application scenario can be represented as a specific desired depth.
Thus, we also take the desired depth as the input of the adapter module, and require the average of predicted depth map to approximate the desired depth.
And the learning of AdaDSR can then be formulated as the joint optimization of reconstruction and network depth loss.
After training, we can dynamically set the desired depth values to accommodate various application scenarios, and then adopt our AdaDSR to meet the efficiency constraints.

Experiments are conducted to assess our AdaDSR.
Without loss of generality, we adopt EDSR~\cite{EDSR} model as the backbone of our AdaDSR (denoted by AdaEDSR).
It can be observed from Fig.~\ref{fig:intro}(b) that the predicted depth map has smaller depth values for the smooth regions and higher ones for the regions with rich small-scale details.
As shown in Fig.~\ref{fig:intro}(d), our AdaDSR can be flexibly tuned to meet various efficiency constraints (\eg, {FLOPs}) by specifying proper desired depth values.
In contrast, most existing SISR methods can only be performed with deterministic inference and fixed computational cost.
Quantitative and qualitative results further show the effectiveness and adaptability of our AdaDSR in comparison to the state-of-the-art deep SISR methods.
Furthermore, we also take another representative SISR model RCAN~\cite{RCAN} as the backbone model (denoted by AdaRCAN), which illustrates the generality of our AdaDSR.
Considering the training efficiency, ablation analyses are performed on AdaDSR with EDSR backbone (\ie, AdaEDSR).

To sum up, the contributions of this work include:
\begin{itemize}
    \item We present adaptive inference networks for deep SISR, \ie, AdaDSR, which adds the backbone with a lightweight adapter module to produce local depth map for spatially adaptive inference.
    \item Both image features and desired depth are taken as the input of the adapter, and reconstruction loss is incorporated with depth loss for network learning, thereby making AdaDSR equipped with sparse convolution to be adaptive to various efficiency constraints.
    \item Experiments show that our AdaDSR achieves better tradeoff between accuracy and efficiency than it counterparts (\ie, EDSR and RCAN), and can adapt to different efficiency constraints without training from scratch.
\end{itemize}

\section{Related Work}
In this section, we briefly review several topics relevant to our AdaDSR, including deep SISR models and adaptive inference methods.

\subsection{Deep Single Image Super-Resolution}
Dong \etal introduce a three-layer convolutional network in their pioneer work SRCNN~\cite{SRCNN}, since then, the quantitative performance of SISR has been continuously promoted with the rapid development of CNNs.
Kim \etal~\cite{VDSR} further propose a deeper model named VDSR with residual blocks and adjustable gradient clipping.
Liu \etal~\cite{MWCNN} propose MWCNN, which accelerates the running speed and enlarges the receptive field by deploying U-Net~\cite{U-Net} like architecture, and multi-scale wavelet transformation is applied rather than traditional down-sampling or up-sampling module to avoid information lost.

These methods take interpolated LR images as input, resulting in heavy computation burden, so many recent SISR methods choose to increase the spatial resolution via PixelShuffle~\cite{PixelShuffle} at the tail of the model.
SRResNet~\cite{SRResNet}, EDSR~\cite{EDSR} and WDSR~\cite{WDSR} follow this setting and have a deep main body by stacking several identical residual blocks~\cite{ResNet} before the tail component, and they obtain better performance and efficiency by modifying the architecture of the residual blocks.
Zhang \etal~\cite{RCAN} build a very deep (more than 400 layers) yet narrow (64 channels \vs 256 channels in EDSR) RCAN model and learn a content-related weight for each feature channel inside the residual blocks.
Dai \etal~\cite{SAN} propose SAN to obtain better feature representation via second-order attention model, and non-locally enhanced residual group is incorporated to capture long-distance features.

Apart from the fidelity track, considerable attention has also been given to handle several other issues in SISR.
For example, SRGAN~\cite{SRResNet} incorporates adversarial loss to improve perceptual quality,
DPSR~\cite{DPSR} proposes a new degradation model and performs super-resolution and deblurring simultaneously,
Zhang \etal~\cite{zhang2019multiple} solve real image SISR problem in an unsupervised manner by taking advantage of generative adversarial networks.
In addition, lightweight networks such as IDN~\cite{IDN} and CARN~\cite{CARN} are proposed, but most lightweight models are accelerated at the cost of quantitative performance.
In this paper, we propose an AdaDSR model, which achieves better tradeoff between accuracy and efficiency.

\subsection{Adaptive Inference}
Traditional deterministic CNNs tend to be less flexible to meet various requirements in the applications.
As a remedy, many adaptive inference methods have been explored in recent years.
Inspired by \cite{bengio2013better}, Upchurch \etal~\cite{upchurch2017deep} propose to learn an interpolation of deep features extracted by a pre-trained model, and manipulate the attributes of facial images.
Shoshan \etal~\cite{DynamicNet} further propose a dynamic model named DynamicNet by deploying tuning blocks alongside the backbone model, and linearly manipulate the features to learn an interpolation of two objectives, which can be tuned to explore the whole objective space during the inference phase.
Similarly, CFSNet~\cite{CFSNet} implements continuous transition of different objectives, and automatically learns the trade-off between perception and distortion for SISR.

Some methods also leverage adaptive inference to obtain computing efficient models.
Li \etal~\cite{li2019improved} deploy multiple classifiers between the main blocks, and the last one performs as a teacher net to guide the previous ones. During the inference phase, the confidence score of a classifier indicates whether to perform the next block and the corresponding classifier.
Figurnov \etal~\cite{patch_adaptive} predict a stop score for the patches, which determines whether to skip the subsequent layers, indicating different regions have unequal importance for detection tasks. Therefore, skipping layers at less important regions can save the inference time.
Yu \etal~\cite{Path-restore} propose to build a denoising model with several multi-path blocks, and in each block, a path finder is deployed to select a proper path for each image patch.
These methods are similar to our AdaDSR, however, they perform adaptive inference on patch-level, and the adaptation depends only on the features.
In this paper, our AdaDSR implements pixel-wise adaptive inference via sparse convolution and is manually controllable to meet various resource constraints.

\section{Proposed Method}
\begin{figure*}[t]
    \centering
    \includegraphics[width=\linewidth]{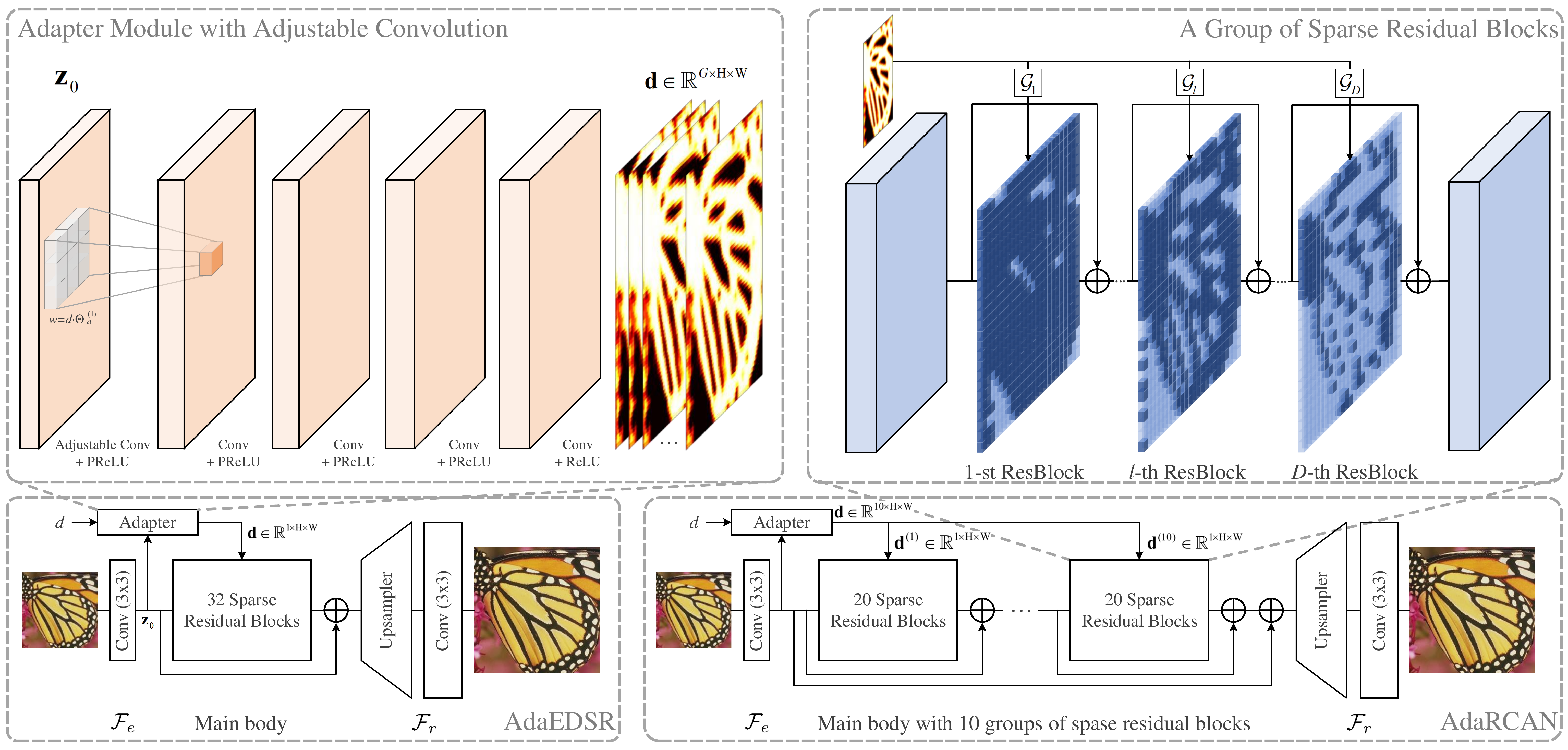}
    \vspace{-2em}
    \caption{\footnotesize Overall illustration of AdaDSR. On the bottom are diagrams showing AdaEDSR and AdaRCAN, respectively. On the top left, a five-layer adapter takes $\mathbf{z}_0$ as input and the weight of the first convolution is tuned by $d$ on the fly. The adapter generats a depth map $\mathbf{d}\in\mathbb{R}^{G\times\mathrm{H}\times\mathrm{W}}$ (for AdaEDSR $G=1$, while for AdaRCAN $G=10$). Each channel of $\mathbf{d}$ is delivered to a group of sparse residual blocks (as shown on the top right). Only a fraction of the positions (marked by dark blue) require computation.}
    \label{fig:network}
    \vspace{-1em}
\end{figure*}

This section presents our AdaDSR model for single image super-resolution.
To begin with, we equip the backbone with a network depth map to facilitate spatially variant inference.
Then, sparse convolution is introduced to speed up the inference by omitting the unnecessary computation.
Furthermore, a lightweight adapter module is deployed to predict the network depth map.
Finally, the overall network structure (see Fig.~\ref{fig:network}) and learning objective are provided.

\subsection{AdaDSR with Spatially Variant Network Depth}
Single image super-resolution aims at learning a mapping to reconstruct the high-resolution image $\hat{\mathbf{y}}$ from its low-resolution (LR) observation $\mathbf{x}$, and can be written as,

\begin{equation}
    \hat{\mathbf{y}} = \mathcal{F}(\mathbf{x}; \mathrm{\Theta}),
    \label{eqn:deepSR}
\end{equation}
where $\mathcal{F}$ denotes the SISR network with the network parameters $\mathrm{\Theta}$.
In this work, we consider a representative category of deep SISR networks that consist of three major modules, \ie, feature extraction $\mathcal{F}_e$, residual blocks, and HR reconstruction $\mathcal{F}_r$.
Several representative SISR models, \eg, SRResNet~\cite{SRResNet}, EDSR~\cite{EDSR}, and RCAN~\cite{RCAN}, belong to this category.
Using EDSR as an example, we let $\mathbf{z}_0 = \mathcal{F}_e(\mathbf{x})$.
The output of the residual blocks can then be formulated as,
\begin{equation}
    \mathbf{z}^{o} = \mathbf{z}_0 + \sum_{l=1}^{D} \mathcal{F}_l(\mathbf{z}_{l-1}; \mathrm{\Theta}_l),
    \label{eqn:ResBlocks}
\end{equation}
where $\mathrm{\Theta}_l$ is the network parameters associated with the $l$-th residual block.
Given the output of the $(l-1)$-th residual block, the $l$-th residual block can be written as $\mathbf{z}_{l} = \mathbf{z}_{l-1} + \mathcal{F}_l(\mathbf{z}_{l-1}; \mathrm{\Theta}_l)$.
Finally, the reconstructed HR image can be obtained by $\hat{\mathbf{y}} = \mathcal{F}_r(\mathbf{z}^{o}; \mathrm{\Theta}_r)$.

As shown in Fig.~\ref{fig:intro}, the difficulty of super-resolution is spatially variant.
For examples, it is not required to go through all the $D$ residual blocks in Eqn.~(\ref{eqn:ResBlocks}) to reconstruct the smooth regions.
As for the regions with rich and detailed textures, more residual blocks generally are required to fulfill high quality reconstruction.
Therefore, we introduce a 2D network depth map $\mathbf{d}$ ($0 \leq d_{ij} \leq D $) which has the same spatial size with $\mathbf{z}_0$.
Intuitively, the network depth $d_{ij}$ is smaller for the smooth region and larger for the region with rich details.
To facilitate spatially adaptive inference, we modify Eqn.~(\ref{eqn:ResBlocks}) as,
\begin{equation}
    \mathbf{z}^{o} = \mathbf{z}_0 + \sum_{l=1}^{D} \mathcal{G}_l(\mathbf{d}) \circ \mathcal{F}_l(\mathbf{z}_{l-1}; \mathrm{\Theta}_l),
    \label{eqn:AdaResBlocks}
\end{equation}
where $\circ$ denotes the entry-wise product.
Here, $\mathcal{G}_l({d}_{ij})$ is defined as,
\begin{equation}
    \mathcal{G}_l(d_{ij})=\left\{
    \begin{array}{cl}
        0,\ \ & d_{ij} < l-1 \\
        1,\ \ & d_{ij} > l \\
        d_{ij}-(l-1),\ \ & otherwise
    \end{array}\right..\label{eqn:MaskFunc}
\end{equation}

Let $\lceil \cdot \rceil$ be the ceiling function,
thus, the last $D - \lceil d_{ij} \rceil$ residual blocks are not required to compute for a position with the network depth $d_{ij}$.
Given the 2D network depth map $\mathbf{d}$, we can then exploit Eqn.~(\ref{eqn:AdaResBlocks}) to conduct spatially adaptive inference.

\begin{figure}[t]
    \centering
    \includegraphics[width=.9\linewidth]{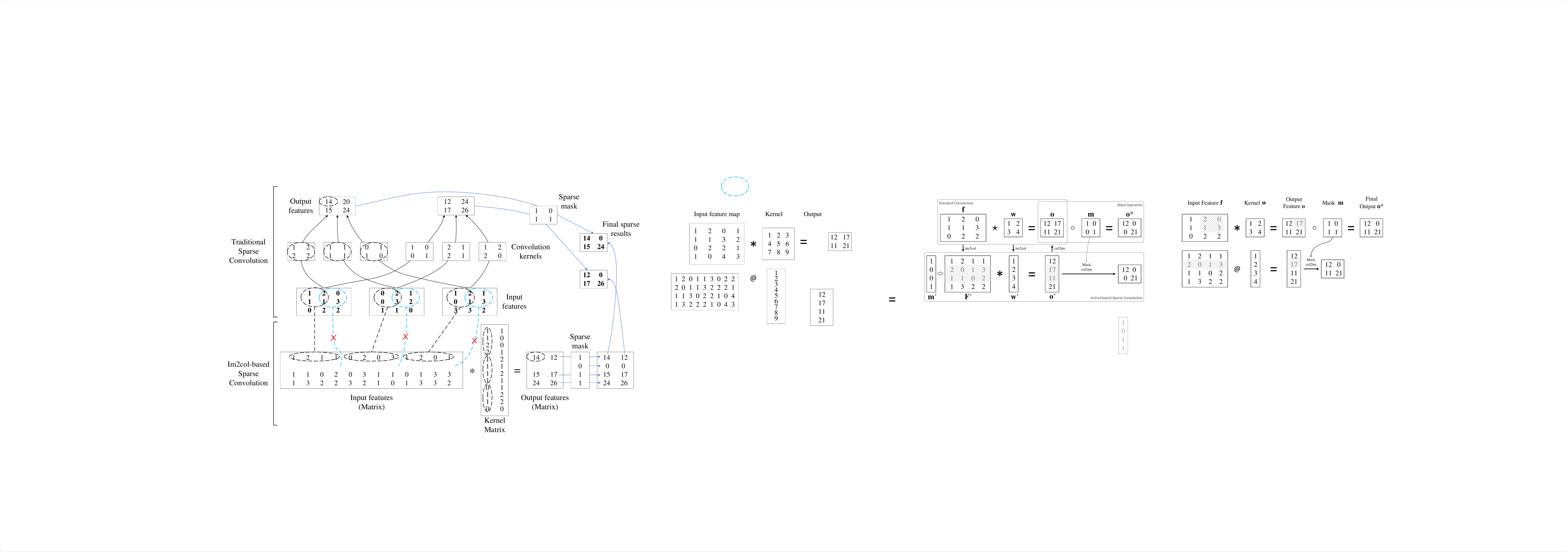}
    \caption{\footnotesize An example to illustrate the im2col~\cite{im2col} based sparse convolution. $\star$, $\circ$ and $\ast$ represent convolution, entry-wise product and matrix multiplication, respectively. $\mathbf{f}$, $\mathbf{w}$ and $\mathbf{o}$ are input feature, convolution kernel and output feature of standard convolution operation, which is implemented by arbitrary convolution implementation algorithms, while $\mathbf{F}'$ and $\mathbf{w}'$ are reorganized from $\mathbf{f}$ and $\mathbf{w}$ during the im2col procedure. Given the mask $\mathbf{m}$, the reorganized $\mathbf{m}'$ indicates that the shaded rows can be safely ignored in the im2col based sparse convolution, therefore reducing computation amount comparing to standard convolution based sparse convolution (as shown in the upper half).}
    \vspace{-0.5em}
    \label{fig:im2col}
    \vspace{-1em}
\end{figure}

\subsection{Sparse Convolution for Efficient Inference}
\vspace{-1em}
Let $\mathbf{m}$ (\eg, $\mathbf{m}_l = \mathcal{G}_l(\mathbf{d})$ for the $l$-th residual block) be a mask to indicate the positions where the convolution activations should be kept.
As shown in Fig.~\ref{fig:im2col}, for some convolution implementations such as fast Fourier transform (FFT)~\cite{fft-1,fft-2} and Winograd~\cite{winograd} based algorithms, one should first perform the standard convolution to obtain the whole output feature map by $\mathbf{o}=\mathbf{w}\star\mathbf{f}$.
Here, $\mathbf{f}$, $\mathbf{w}$ and $\star$ denote input feature map, convolution kernel and convolution operation, respectively.
Then the sparse results can be represented by $\mathbf{o}^{*} = \mathbf{m} \circ \mathbf{o}$.
Nonetheless, such implementations meet the requirement of spatially adaptive inference while maintaining the same computational complexity with the standard convolution.

Instead, we adopt the im2col~\cite{im2col} based sparse convolution for efficient adaptive inference.
As shown in Fig.~\ref{fig:im2col}, the patch from $\mathbf{f}$ related to a point in $\mathbf{o}$ is organized as a row in matrix $\mathbf{F}'$, and the convolution kernel ($\mathbf{w}$) is also converted as vector $\mathbf{w}'$.
Then the convolution operation is transformed into a matrix multiplication problem, which is highly optimized in many Basic Linear Algebra Subprograms (BLAS) libraries.
Then, the result $\mathbf{o}'$ can be organized back to the output feature map.
Given the mask $\mathbf{m}$, we can simply skip the corresponding row when constructing the reorganized input feature $\mathbf{F}'$ if it has zero mask value (see the shaded rows of $\mathbf{F}'$ in Fig.~\ref{fig:im2col}), and the computation is skipped as well.
Thus, the spatially adaptive inference in Eqn.~(\ref{eqn:AdaResBlocks}) can be efficiently implemented via the im2col and col2im procedure.
Moreover, the efficiency can be further improved when more rows are masked out, \ie, when the average depth of $\mathbf{d}$ is smaller.

It is worth noting that sparse convolution has been suggested in many works and evaluated in image classification~\cite{SparseWinogradConv}, object detection~\cite{SparseCNN,SBNet}, model pruning~\cite{FasterCNN} and 3D semantic segmentation~\cite{SparseConvNet} tasks.
\cite{SparseCNN} and \cite{SBNet} are based on im2col and Winograd algorithm respectively, however, these methods implement patch-level sparse convolution.
\cite{SparseConvNet} designs new data structure for sparse convolution and constructs a whole CNN framework to suit the designed data structure, making it incompatible with standard methods.
\cite{SparseWinogradConv} incorporates sparsity into Winograd algorithm, which is not mathematically equivalent to the vanilla CNN nor the conventional Winograd CNN.
The most relevant work \cite{FasterCNN} skips unnecessary points when traversing all spatial positions and achieves pixel-level sparse convolution, which is implemented on serial devices (\eg, CPUs) via for-loops.
In this work, we use im2col based sparse convolution, which combines this intuitive thought and im2col algorithm, and deploy the proposed model on the parallel platforms (\eg, GPUs).
To the best of our knowledge, this is the first attempt to deploy pixel-wise sparse convolution on SISR task and achieves image content and resource adaptive inference.

\vspace{-1em}
\subsection{Lightweight Adapter Module}
In this subsection, we introduce a lightweight adapter module $\mathcal{P}$ to predict a 2D network depth map $\mathbf{d}$.
In order to adapt to local image content, the adapter module $\mathcal{P}$ is required to produce lower network depth for smooth region and higher depth for detailed region.
Let $\bar{d}$ be the average value of $\mathbf{d}$, and $d$ be the desired network depth.
To make the model to be adaptive to efficiency constraints, we also take the desired network depth $d$ into account, and require that the decrease of $d$ can result in smaller $\bar{d}$, \ie, better inference efficiency.

As shown in Fig.~\ref{fig:network}, the adapter module $\mathcal{P}$ takes the feature map $\mathbf{z}_0$ as the input and is comprised of four convolution layers with PReLU nonlinearity followed by another convolution layer with ReLU nonlinearity.
Let $\mathbf{d} = \mathcal{P}(\mathbf{z}_0; \mathrm{\Theta}_a)$.
We then use Eqn.~(\ref{eqn:MaskFunc}) to generate the mask $\mathbf{m}_l$ for each residual block.
It is noted that $\mathbf{m}_l$ may not be a binary mask but contains many zeros.
Thus, we can construct a sparse residual block which can omit the computation for the regions with zero mask values to facilitate efficient adaptive inference.
To meet the efficiency constraint, we also take the desired network depth $d$ as the input to the adapter, and predict the network depth map by
\begin{equation}
    \mathbf{d} = \mathcal{P}(\mathbf{z}_0, d; \mathrm{\Theta}_a),
\end{equation}
where $\mathrm{\Theta}_a$ denotes the network parameters of the adapter module.
Specifically, denote the weight of the first convolution layer in the adapter as $\mathrm{\Theta}^{(1)}_a$,
we make the convolution adjustable by replacing the weight $\mathrm{\Theta}^{(1)}_a$ with $d\cdot\mathrm{\Theta}^{(1)}_a$ when the desired depth is $d$, therefore the adapter is able to meet the aforementioned $d$-oriented constraints.

\subsection{Network Architecture and Learning Objective}

\noindent\textbf{Network Architecture}.
%
As shown in Fig.~\ref{fig:network}, our proposed AdaDSR is comprised of a backbone SISR network and a lightweight adapter module to facilitate image content and efficiency adaptive inference.
Without loss of generality, in this section, we take EDSR~\cite{EDSR} as the backbone to illustrate the network architecture, and it is feasible to apply our AdaDSR to other representative SISR models~\cite{SRResNet, WDSR, RCAN} with a number of residual blocks~\cite{ResNet}.
Following~\cite{EDSR}, the backbone involves 32 residual blocks, each of which has two $3\times3$ convolution layers with stride 1, padding 1 and 256 channels with ReLU nonlinearity. Another $3\times3$ convolution layer is deployed right behind the residual blocks.
The feature extraction module $\mathcal{F}_e$ is a convolution layer,
and the reconstruction module $\mathcal{F}_r$ is comprised of an upsampling unit to enlarge the features followed by a convolution layer which reconstructs the output image.
The upsampling unit is composed by a series of Convolution-PixelShuffle~\cite{PixelShuffle} according to the super-resolution scale.
Besides, the lightweight adapter module takes both the feature map $\mathbf{z}_0$ and the desired network depth $d$ as the input, and consists of five convolution layers to produce an one-channel network depth map.

It is worth noting that, we implement two versions of AdaDSR.
The first takes EDSR~\cite{EDSR} as backbone, which is denoted by AdaEDSR.
To further show the generality of proposed AdaDSR and compare against state-of-the-art methods, we also take RCAN~\cite{RCAN} as backbone and implement an AdaRCAN model.
The main difference is that, RCAN replaces the 32 residual blocks with 10 residual groups, and each residual groups is composed of 20 residual blocks equipped with channel attention.
Therefore, we modify the adapter to generate 10 depth maps simultaneously, and each of which is deployed to a residual group.

\vspace{1em}
\noindent\textbf{Learning Objective}.
The learning objective of our AdaDSR includes a reconstruction loss term and a network depth loss term to achieve a proper tradeoff between SISR performance and efficiency.
In terms of the SISR performance, we adopt the $\ell_1$ reconstruction loss defined on the super-resolved output and the ground-truth high-resolution image,
\begin{equation}
        \mathcal{L}_\mathit{rec}\
             =\|\mathbf{y} - \hat{\mathbf{y}}\|_1,
\end{equation}
where $\mathbf{y}$ and $\hat{\mathbf{y}}$ respectively represent the high-resolution ground-truth and the super-resolved image by our AdaDSR.
Considering the efficiency constraint, we require the average $\bar{d}$ of the predicted network depth map to approximate the desired depth $d$, and then introduce the following network depth loss,
\begin{equation}
    \mathcal{L}_\mathit{depth} = \max(0, \bar{d} - d).
\end{equation}
To sum up, the overall learning objective of our AdaDSR is formulated as,
\begin{equation}
    \mathcal{L} = \mathcal{L}_\mathit{rec} + \lambda \mathcal{L}_\mathit{depth},
\end{equation}
where $\lambda$ is a tradeoff hyper-parameter and is set to $0.01$ in all our experiments.

\section{Experiments}\label{sec:experiments}
\begin{table*}[t]
    \center
    \begin{center}
    \caption{\footnotesize Quantitative results in comparison with the state-of-the-art methods. Best three methods are highlighted by {\color{red}red}, {\color{blue}blue} and {\color{green}green}, respectively.}
    \label{tab:performance}
    \scalebox{0.9} {
    \begin{tabular}{|l|c|c|c|c|c|c|c|c|c|c|c|}
    \hline
    \multirow{2}{*}{Method} & \multirow{2}{*}{Scale} &  \multicolumn{2}{c|}{Set5} &  \multicolumn{2}{c|}{Set14} &  \multicolumn{2}{c|}{B100} &  \multicolumn{2}{c|}{Urban100} &  \multicolumn{2}{c|}{Manga109}
    \\
    \cline{3-12}
    &  & PSNR & SSIM & PSNR & SSIM & PSNR & SSIM & PSNR & SSIM & PSNR & SSIM
    \\
    \hline
    \hline

Bicubic & $\times2$
   & 33.66 & 0.9299   
   & 30.24 & 0.8688   
   & 29.56 & 0.8431   
   & 26.88 & 0.8403   
   & 30.80 & 0.9339   
\\

SRCNN~\cite{SRCNN} & $\times2$
   & 36.66 & 0.9542   
   & 32.45 & 0.9067   
   & 31.36 & 0.8879   
   & 29.50 & 0.8946   
   & 35.60 & 0.9663   
\\

VDSR~\cite{VDSR} & $\times2$
   & 37.53 & 0.9590   
   & 33.05 & 0.9130   
   & 31.90 & 0.8960   
   & 30.77 & 0.9140   
   & 37.22 & 0.9750   
\\

EDSR~\cite{EDSR} & $\times2$
   & 38.11 & 0.9602   
   & 33.92 & 0.9195   
   & 32.32 & 0.9013   
   & 32.93 & 0.9351   
   & 39.10 & 0.9773   
\\

AdaEDSR & $\times2$
   & 38.21 & 0.9611   
   & 33.97 & 0.9208   
   & 32.35 & 0.9017   
   & 32.91 & 0.9353   
   & 39.11 & 0.9778   
\\

RDN~\cite{RDN} & $\times2$
   & 38.24 & 0.9614   
   & 34.01 & 0.9212   
   & 32.34 & 0.9017   
   & 32.89 & 0.9353   
   & 39.18 & 0.9780   
\\

RCAN~\cite{RCAN} & $\times2$
   & {\color{green}38.27} & {\color{green}0.9614}   
   & {\color{red}34.12} & {\color{red}0.9216}   
   & {\color{blue}32.41} & {\color{blue}0.9027}   
   & {\color{red}33.34} & {\color{red}0.9384}   
   & {\color{red}39.44} & {\color{blue}0.9786}   
\\

SAN~\cite{SAN} & $\times2$
   & {\color{red}38.31} & {\color{red}0.9620}   
   & {\color{green}34.07} & {\color{green}0.9213}   
   & {\color{red}32.42} & {\color{red}0.9028}   
   & {\color{green}33.10} & {\color{green}0.9370}   
   & {\color{green}39.32} & {\color{red}0.9792}   
\\

AdaRCAN & $\times2$
   & {\color{blue}38.28} & {\color{blue}0.9615}   
   & {\color{red}34.12} & {\color{red}0.9216}   
   & {\color{blue}32.41} & {\color{green}0.9026}   
   & {\color{blue}33.29} & {\color{blue}0.9380}   
   & {\color{red}39.44} & {\color{green}0.9785}   
\\
\hline
\hline

Bicubic & $\times3$
   & 30.39 & 0.8682   
   & 27.55 & 0.7742   
   & 27.21 & 0.7385   
   & 24.46 & 0.7349   
   & 26.95 & 0.8556   
\\

SRCNN~\cite{SRCNN} & $\times3$
   & 32.75 & 0.9090   
   & 29.30 & 0.8215   
   & 28.41 & 0.7863   
   & 26.24 & 0.7989   
   & 30.48 & 0.9117   
\\

VDSR~\cite{VDSR} & $\times3$
   & 33.67 & 0.9210   
   & 29.78 & 0.8320   
   & 28.83 & 0.7990   
   & 27.14 & 0.8290   
   & 32.01 & 0.9340   
\\

EDSR~\cite{EDSR} & $\times3$
   & 34.65 & 0.9280   
   & 30.52 & 0.8462   
   & 29.25 & 0.8093   
   & 28.80 & 0.8653   
   & 34.17 & 0.9476   
\\

AdaEDSR & $\times3$
   & 34.65 & 0.9288   
   & 30.57 & 0.8463   
   & 29.27 & 0.8091   
   & 28.78 & 0.8649   
   & 34.16 & 0.9482   
\\

RDN~\cite{RDN} & $\times3$
   & 34.71 & 0.9296   
   & 30.57 & 0.8468   
   & 29.26 & 0.8093   
   & 28.80 & 0.8653   
   & 34.13 & 0.9484   
\\

RCAN~\cite{RCAN} & $\times3$
   & {\color{green}34.74} & {\color{green}0.9299}   
   & {\color{red}30.65} & {\color{red}0.8482}   
   & {\color{green}29.32} & {\color{blue}0.8111}   
   & {\color{red}29.09} & {\color{red}0.8702}   
   & {\color{blue}34.44} & {\color{red}0.9499}   
\\

SAN~\cite{SAN} & $\times3$
   & {\color{blue}34.75} & {\color{blue}0.9300}   
   & {\color{green}30.59} & {\color{green}0.8476}   
   & {\color{red}29.33} & {\color{red}0.8112}   
   & {\color{green}28.93} & {\color{green}0.8671}   
   & {\color{green}34.30} & {\color{green}0.9494}   
\\

AdaRCAN & $\times3$
   & {\color{red}34.79} & {\color{red}0.9302}   
   & {\color{red}30.65} & {\color{blue}0.8481}   
   & {\color{red}29.33} & {\color{blue}0.8111}   
   & {\color{blue}29.03} & {\color{blue}0.8689}   
   & {\color{red}34.49} & {\color{blue}0.9498}   
\\
\hline
\hline

Bicubic & $\times4$
   & 28.42 & 0.8104   
   & 26.00 & 0.7027   
   & 25.96 & 0.6675   
   & 23.14 & 0.6577   
   & 24.89 & 0.7866   
\\

SRCNN~\cite{SRCNN} & $\times4$
   & 30.48 & 0.8628   
   & 27.50 & 0.7513   
   & 26.90 & 0.7101   
   & 24.52 & 0.7221   
   & 27.58 & 0.8555   
\\

VDSR~\cite{VDSR} & $\times4$
   & 31.35 & 0.8830   
   & 28.02 & 0.7680   
   & 27.29 & 0.0726   
   & 25.18 & 0.7540   
   & 28.83 & 0.8870   
\\

EDSR~\cite{EDSR} & $\times4$
   & 32.46 & 0.8968   
   & 28.80 & 0.7876   
   & 27.71 & 0.7420   
   & 26.64 & 0.8033   
   & 31.02 & 0.9148   
\\

AdaEDSR & $\times4$
   & 32.49 & 0.8977   
   & 28.76 & 0.7865   
   & 27.71 & 0.7410   
   & 26.58 & 0.8011   
   & 30.96 & 0.9150   
\\

RDN~\cite{RDN} & $\times4$
   & 32.47 & 0.8990   
   & 28.81 & 0.7871   
   & 27.72 & 0.7419   
   & 26.61 & 0.8028   
   & 31.00 & 0.9151   
\\

RCAN~\cite{RCAN} & $\times4$
   & {\color{blue}32.63} & {\color{blue}0.9002}   
   & {\color{green}28.87} & {\color{red}0.7889}   
   & {\color{blue}27.77} & {\color{red}0.7436}   
   & {\color{red}26.82} & {\color{red}0.8087}   
   & {\color{red}31.22} & {\color{red}0.9173}   
\\

SAN~\cite{SAN} & $\times4$
   & {\color{red}32.64} & {\color{red}0.9003}   
   & {\color{red}28.92} & {\color{blue}0.7888}   
   & {\color{red}27.78} & {\color{red}0.7436}   
   & {\color{green}26.79} & {\color{blue}0.8068}   
   & {\color{green}31.18} & {\color{green}0.9169}   
\\

AdaRCAN & $\times4$
   & {\color{green}32.61} & {\color{green}0.8998}   
   & {\color{blue}28.88} & {\color{green}0.7883}   
   & {\color{blue}27.77} & {\color{green}0.7428}   
   & {\color{blue}26.80} & {\color{green}0.8067}   
   & {\color{red}31.22} & {\color{blue}0.9172}   
\\

    \hline
    \end{tabular} }
    \end{center}
    \vspace{-1em}
\end{table*}

\begin{table*}[t]
    \center
    \begin{center}
    \caption{\footnotesize Inference efficiency in comparison with the state-of-the-art methods. Note that the GPU memory is not enough to run SAN~\cite{SAN} with scale $\times2$ on Urban100 and Manga109 datasets.}
    \label{tab:efficiency}
    \scalebox{0.84} {
    \begin{tabular}{|l|c|c|c|c|c|c|c|c|c|c|c|}
    \hline
    \multirow{3}{*}{Method} & \multirow{3}{*}{Scale} &  \multicolumn{2}{c|}{Set5} &  \multicolumn{2}{c|}{Set14} &  \multicolumn{2}{c|}{B100} &  \multicolumn{2}{c|}{Urban100} &  \multicolumn{2}{c|}{Manga109}
    \\
    \cline{3-12}
    &  & FLOPs & Time & FLOPs & Time & FLOPs & Time & FLOPs & Time & FLOPs & Time \\
    &  & (G) & (ms) & (G) & (ms) & (G) & (ms) & (G) & (ms) & (G) & (ms)
    \\
    \hline
    \hline

SRCNN~\cite{SRCNN} & $\times2$
   & 6.1 & 43.0   
   & 12.3 & 7.9   
   & 8.2 & 4.4   
   & 41.4 & 19.2   
   & 51.6 & 24.2   
\\

VDSR~\cite{VDSR} & $\times2$
   & 70.5 & 86.9   
   & 143.0 & 88.7   
   & 95.4 & 58.0   
   & 481.6 & 301.0   
   & 599.5 & 368.8   
\\

EDSR~\cite{EDSR} & $\times2$
   & 1338.8 & 395.7   
   & 2552.2 & 630.7   
   & 1776.9 & 469.9   
   & 8041.1 & 2163.8   
   & 9891.4 & 2554.5   
\\

AdaEDSR & $\times2$
   & 650.6 & 312.3   
   & 1397.3 & 489.8   
   & 965.3 & 371.5   
   & 4844.9 & 1655.2   
   & 5208.4 & 1864.5   
\\

RDN~\cite{RDN} & $\times2$
   & 801.1 & 345.9   
   & 1527.3 & 617.1   
   & 1063.3 & 407.7   
   & 4811.9 & 2198.5   
   & 5919.2 & 3417.1   
\\

RCAN~\cite{RCAN} & $\times2$
   & 577.9 & 633.2   
   & 1101.8 & 813.6   
   & 767.0 & 607.0   
   & 3471.2 & 1955.0   
   & 4270.0 & 2342.9   
\\

SAN~\cite{SAN} & $\times2$
   & 3835.9 & 1276.0   
   & 17500.4 & 3314.0   
   & 3943.2 & 1637.8   
   & 372727.5 & N/A    
   & 645359.4 & N/A    
\\

AdaRCAN & $\times2$
   & 469.1 & 614.5   
   & 925.5 & 751.9   
   & 649.2 & 606.3   
   & 2907.2 & 1749.2   
   & 3300.7 & 2034.2   
\\
\hline
\hline

SRCNN~\cite{SRCNN} & $\times3$
   & 6.1 & 43.0   
   & 12.3 & 7.9   
   & 8.2 & 4.4   
   & 41.4 & 19.2   
   & 51.6 & 24.2   
\\

VDSR~\cite{VDSR} & $\times3$
   & 70.5 & 86.9   
   & 143.0 & 88.7   
   & 95.4 & 58.0   
   & 481.6 & 301.0   
   & 599.5 & 368.8   
\\

EDSR~\cite{EDSR} & $\times3$
   & 699.1 & 251.0   
   & 1305.7 & 341.7   
   & 924.1 & 259.6   
   & 3984.0 & 957.9   
   & 4904.0 & 1271.1   
\\

AdaEDSR & $\times3$
   & 504.8 & 231.4   
   & 1013.5 & 302.0   
   & 722.8 & 232.0   
   & 3314.2 & 858.4   
   & 3695.9 & 1023.3   
\\

RDN~\cite{RDN} & $\times3$
   & 437.1 & 195.0   
   & 816.3 & 290.3   
   & 577.8 & 190.2   
   & 2490.9 & 1045.3   
   & 3066.1 & 1404.0   
\\

RCAN~\cite{RCAN} & $\times3$
   & 328.5 & 553.6   
   & 613.5 & 551.7   
   & 434.2 & 511.2   
   & 1872.0 & 1029.2   
   & 2304.2 & 1208.7   
\\

SAN~\cite{SAN} & $\times3$
   & 463.2 & 582.2   
   & 1930.1 & 992.9   
   & 517.5 & 600.7   
   & 36735.2 & 5416.2   
   & 61976.0 & 8194.4   
\\

AdaRCAN & $\times3$
   & 277.7 & 572.6   
   & 512.9 & 559.1   
   & 369.3 & 523.8   
   & 1596.3 & 968.1   
   & 1842.2 & 1107.1   
\\
\hline
\hline

SRCNN~\cite{SRCNN} & $\times4$
   & 6.1 & 43.0   
   & 12.3 & 7.9   
   & 8.2 & 4.4   
   & 41.4 & 19.2   
   & 51.6 & 24.2   
\\

VDSR~\cite{VDSR} & $\times4$
   & 70.5 & 86.9   
   & 143.0 & 88.7   
   & 95.4 & 58.0   
   & 481.6 & 301.0   
   & 599.5 & 368.8   
\\

EDSR~\cite{EDSR} & $\times4$
   & 501.9 & 214.6   
   & 908.8 & 239.8   
   & 655.7 & 240.8   
   & 2699.4 & 640.0   
   & 3297.6 & 762.2   
\\

AdaEDSR & $\times4$
   & 371.7 & 181.1   
   & 716.8 & 215.4   
   & 508.5 & 195.2   
   & 2265.8 & 563.3   
   & 2588.4 & 656.0   
\\

RDN~\cite{RDN} & $\times4$
   & 337.9 & 128.3   
   & 611.9 & 163.7   
   & 441.5 & 132.9   
   & 1817.3 & 512.4   
   & 2220.9 & 646.2   
\\

RCAN~\cite{RCAN} & $\times4$
   & 270.1 & 546.9   
   & 489.0 & 505.7   
   & 352.8 & 490.0   
   & 1452.5 & 684.7   
   & 1774.4 & 843.8   
\\

SAN~\cite{SAN} & $\times4$
   & 159.4 & 482.7   
   & 522.2 & 568.5   
   & 190.9 & 445.2   
   & 7770.0 & 2258.0   
   & 12858.7 & 3174.3   
\\

AdaRCAN & $\times4$
   & 227.5 & 561.6   
   & 418.1 & 524.9   
   & 304.8 & 520.0   
   & 1263.0 & 659.8   
   & 1463.3 & 712.8   
\\

    \hline
    \end{tabular} }
    \end{center}
    \vspace{-1em}
\end{table*}

\begin{figure}[t]
    \newlength\fsdurthree
    \setlength{\fsdurthree}{-1.5mm}
    \scriptsize
    \centering
    \begin{tabular}{cc}
    \begin{adjustbox}{valign=t}
\tiny
    \begin{tabular}{c}
        \includegraphics[width=0.21\textwidth, height=\widthscalefive \textwidth]{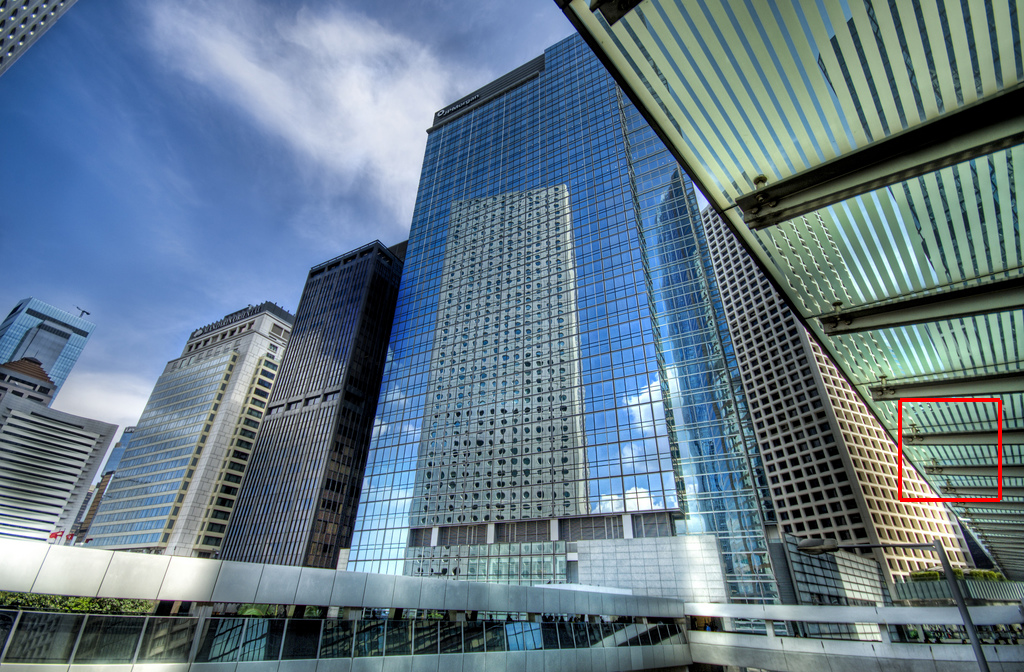}
        \\
        Urban100 ($\times 4$):
        \\
        img\_061
        \\
        \includegraphics[width=0.21\textwidth, height=\widthscalefive \textwidth]{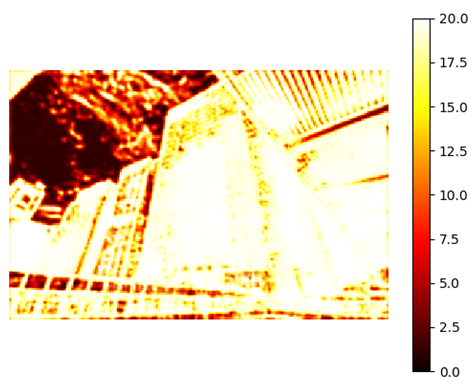}
        \\
        AdaRCAN
        \\
        depth map $\mathbf{d}$
    \end{tabular}
\end{adjustbox}
\hspace{-2.3mm}
\begin{adjustbox}{valign=t}
\tiny
    \begin{tabular}{cccccc}
        \includegraphics[width=\widthscalefive \textwidth]{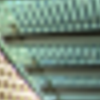} \hspace{\fsdurthree} &
        \includegraphics[width=\widthscalefive \textwidth]{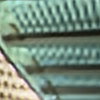} \hspace{\fsdurthree} &
        \includegraphics[width=\widthscalefive \textwidth]{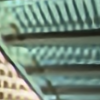} \hspace{\fsdurthree} &
        \includegraphics[width=\widthscalefive \textwidth]{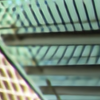} \hspace{\fsdurthree} &
        \includegraphics[width=\widthscalefive \textwidth]{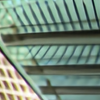} &
        \\
        Bicubic \hspace{\fsdurthree} &
        SRCNN~\cite{SRCNN} \hspace{\fsdurthree} &
        VDSR~\cite{VDSR} \hspace{\fsdurthree} &
        EDSR~\cite{EDSR} \hspace{\fsdurthree} &
        AdaEDSR
        \\
        19.22/0.4316 \hspace{\fsdurthree} &
        19.97/0.5558 \hspace{\fsdurthree} &
        20.16/0.5748 \hspace{\fsdurthree} &
        20.96/0.6616 \hspace{\fsdurthree} &
        20.91/0.6642 
        \\
        \includegraphics[width=\widthscalefive \textwidth]{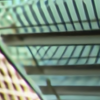} \hspace{\fsdurthree} &
        \includegraphics[width=\widthscalefive \textwidth]{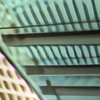} \hspace{\fsdurthree} &
        \includegraphics[width=\widthscalefive \textwidth]{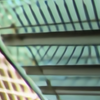} \hspace{\fsdurthree} &
        \includegraphics[width=\widthscalefive \textwidth]{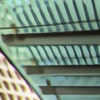} \hspace{\fsdurthree} &
        \includegraphics[width=\widthscalefive \textwidth]{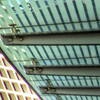}  
        \\ 
        RDN~\cite{RDN} \hspace{\fsdurthree} &
        RCAN~\cite{RCAN} \hspace{\fsdurthree} &
        SAN~\cite{SAN} \hspace{\fsdurthree} &
        AdaRCAN \hspace{\fsdurthree} &
        HR
        \\
        21.06/0.6752 \hspace{\fsdurthree} &
        22.13/0.7378 \hspace{\fsdurthree} &
        21.67/0.7044 \hspace{\fsdurthree} &
        22.21/0.7482 \hspace{\fsdurthree} &
        PSNR/SSIM \hspace{\fsdurthree} 
        \\
    \end{tabular}
\end{adjustbox}
\vspace{0.5mm}
\\
\begin{adjustbox}{valign=t}
\tiny
    \begin{tabular}{c}
        \includegraphics[width=0.21\textwidth, height=\widthscalefive \textwidth]{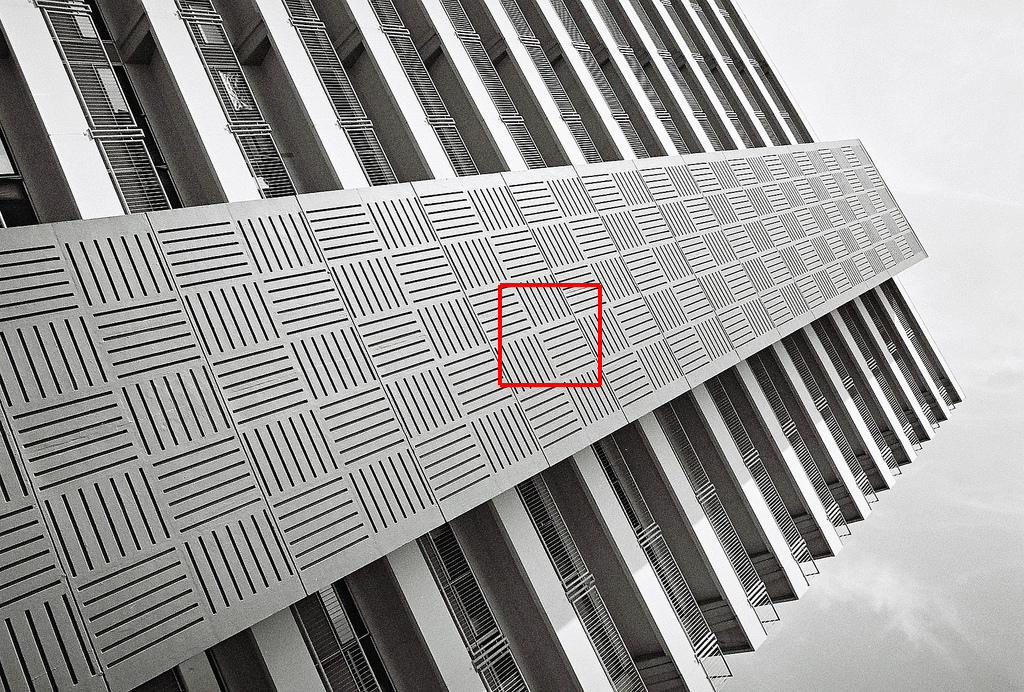}
        \\
        Urban100 ($\times 4$):
        \\
        img\_092
        \\
        \includegraphics[width=0.21\textwidth, height=\widthscalefive \textwidth]{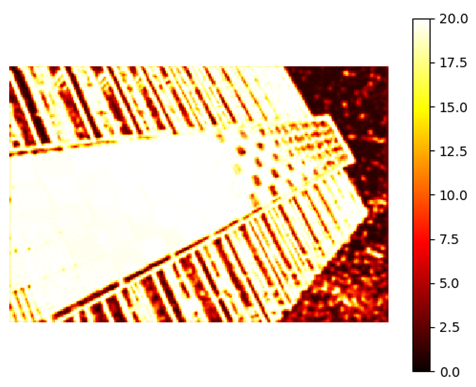}
        \\
        AdaRCAN
        \\
        depth map $\mathbf{d}$
    \end{tabular}
\end{adjustbox}
\hspace{-2.3mm}
\begin{adjustbox}{valign=t}
\tiny
    \begin{tabular}{cccccc}
        \includegraphics[width=\widthscalefive \textwidth]{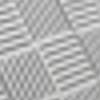} \hspace{\fsdurthree} &
        \includegraphics[width=\widthscalefive \textwidth]{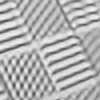} \hspace{\fsdurthree} &
        \includegraphics[width=\widthscalefive \textwidth]{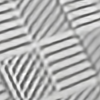} \hspace{\fsdurthree} &
        \includegraphics[width=\widthscalefive \textwidth]{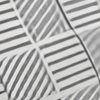} \hspace{\fsdurthree} &
        \includegraphics[width=\widthscalefive \textwidth]{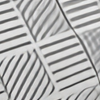} &
        \\
        Bicubic \hspace{\fsdurthree} &
        SRCNN~\cite{SRCNN} \hspace{\fsdurthree} &
        VDSR~\cite{VDSR} \hspace{\fsdurthree} &
        EDSR~\cite{EDSR} \hspace{\fsdurthree} &
        AdaEDSR
        \\
        12.81/0.1879 \hspace{\fsdurthree} &
        13.36/0.3607 \hspace{\fsdurthree} &
        13.13/0.3580 \hspace{\fsdurthree} &
        12.69/0.3297 \hspace{\fsdurthree} &
        12.73/0.3822 
        \\
        \includegraphics[width=\widthscalefive \textwidth]{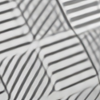} \hspace{\fsdurthree} &
        \includegraphics[width=\widthscalefive \textwidth]{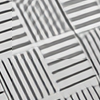} \hspace{\fsdurthree} &
        \includegraphics[width=\widthscalefive \textwidth]{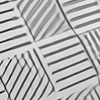} \hspace{\fsdurthree} &
        \includegraphics[width=\widthscalefive \textwidth]{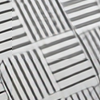} \hspace{\fsdurthree} &
        \includegraphics[width=\widthscalefive \textwidth]{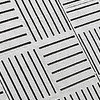}  
        \\ 
        RDN~\cite{RDN} \hspace{\fsdurthree} &
        RCAN~\cite{RCAN} \hspace{\fsdurthree} &
        SAN~\cite{SAN} \hspace{\fsdurthree} &
        AdaRCAN \hspace{\fsdurthree} &
        HR
        \\
        12.92/0.3728 \hspace{\fsdurthree} &
        15.39/0.6261 \hspace{\fsdurthree} &
        13.20/0.4616 \hspace{\fsdurthree} &
        15.53/0.6393 \hspace{\fsdurthree} &
        PSNR/SSIM \hspace{\fsdurthree} 
        \\
    \end{tabular}
\end{adjustbox}

    \vspace{-3mm}
    \end{tabular}
    \caption{\footnotesize
        Visual comparison for $4\times$ SR on Urban100 dataset. Note that the depth map of AdaRCAN is an average of the 10 groups. Kindly refer to the supplementary materials for more results.
    }
    \label{fig:compare}
    \vspace{-3em}
\end{figure}

\subsection{Implementation Details}
\noindent\textbf{Model Training.}
For training our AdaDSR model, we use the 800 training images and the first five validation images from DIV2K dataset~\cite{div2k} as training and validation set, respectively.
The input and output images are in RGB color space, and the input images are obtained by bicubic degradation model.
Following previous works~\cite{SRResNet, EDSR, RCAN}, during training we subtract the mean value of the DIV2K dataset on RGB channels and apply data augmentation on training images, including random horizontal flip, random vertical flip and $90^{\circ}$ rotation.
The AdaDSR model is optimized by the Adam~\cite{adam} algorithm with $\beta_1=0.9$ and $\beta_2=0.999$ for 800 epochs. In each iteration, there are 16 LR patches of size $48\times48$.
And the learning rate is initialized as $5\times10^{-5}$ and decays to half after every 200 epochs.
During training, the desired depth $d$ is randomly sampled from $[0, D]$, where $D$ is 32 and 20 for AdaEDSR and AdaRCAN, respectively.
Note that, due to the data structure of the sparse convolution is identical with standard convolution, we can use the pretrained backbone model to initialize the AdaDSR model to improve the training stability and save training time.

\vspace{0.5em}
\noindent\textbf{Model Evaluation.}
Following previous works~\cite{SRResNet,EDSR,RCAN}, we use PSNR and SSIM as model evaluation metrics, and five standard benchmark datasets (\ie, Set5~\cite{set5}, Set14~\cite{set14}, B100~\cite{b100}, Urban100~\cite{urban100} and Manga109~\cite{manga109}) are employed as test sets, and the PSNR and SSIM indices are calculated on the luminance channel (a.k.a. Y channel) of YCbCr color space with \emph{scale} pixels on the boundary ignored.
Furthermore, the computation efficiency is evaluated by FLOPs and inference time.
For a fair comparison with the competing methods, when counting the running time, we implement all competing methods in our framework and replace the convolution layers of the main body with im2col~\cite{im2col} based convolutions.
All evaluations are conducted in the PyTorch~\cite{pytorch} environment running on a single Nvidia TITAN RTX GPU.
The source code and pre-trained models are publicly available at \href{https://github.com/csmliu/AdaDSR}{\color{red}https://github.com/csmliu/AdaDSR}.
\vspace{-.5em}
\subsection{Comparison with State-of-the-arts}
\vspace{-.5em}
To evaluate the effectiveness of our AdaDSR model, we first compare AdaDSR\footnote{Note that the desired depth $d$ is set to 32 and 20 for AdaEDSR and AdaRCAN in Tables~\ref{tab:performance} and \ref{tab:efficiency} respectively, \ie, the number of residual blocks in EDSR and that of each group in RCAN.} with the backbone EDSR~\cite{EDSR} and RCAN~\cite{RCAN} models as well as four other state-of-the-art methods, \ie, SRCNN~\cite{SRCNN}, VDSR~\cite{VDSR}, RDN~\cite{RDN} and SAN~\cite{SAN}.
Note that all visual results of other methods given in this section are generated by the officially released models, while the FLOPs and inference time are evaluated in our framework.

As shown in Table~\ref{tab:performance}, both AdaEDSR and AdaRCAN perform favorably against their counterparts EDSR and RCAN in terms of quantitative PSNR and SSIM metrics.
Besides, it can be seen from Table~\ref{tab:efficiency}, although the adapter module introduces extra computation cost, it is very lightweight and efficient in comparison to the backbone super-resolution model, and the deployment of the lightweight adapter module greatly reduces computation amount of the whole model, resulting in lower FLOPs and faster inference, especially on large images (\eg, Urban100 and Manga109).
Note that SAN~\cite{SAN} has similar performance with RCAN and AdaRCAN, yet its computation cost is too heavy on large images.

Apart from the quantitative comparison, visual results are also given in Fig.~\ref{fig:compare}.
One can see that AdaEDSR and AdaRCAN are able to generate super-resolved images of similar or better visual quality to their counterparts.
Kindly refer to the supplementary materials for more qualitative results.
We also show the pixel-wise depth map $\mathbf{d}$ of AdaRCAN (due to space limit, we show the average of the depth maps for 10 groups of AdaRCAN) to discuss the relationship between the processed image and the depth map.
As we can see from Fig.~\ref{fig:compare}, greater depth is predicted for the regions with detailed textures, while most of the computation in smooth areas can be omitted for efficiency purpose, which is intuitive and verifies our discussions in Sec.~\ref{sec:intro}.

Considering both quantitative and qualitative results, our AdaDSR can achieve comparable performance against state-of-the-art methods while greatly reducing the computation amount.
Further analysis on the adaptive adjustment of $d$ please refer to Sec.~\ref{sec:adaptive_inference}.

\vspace{-1.5em}
\subsection{Adaptive Inference with Varying Depth}\label{sec:adaptive_inference}
\vspace{-0.5em}

\begin{figure}[t]
    \subfigure[Scale $\times2$]{
    \begin{minipage}{0.3\linewidth}
            \includegraphics[width=.95\linewidth]{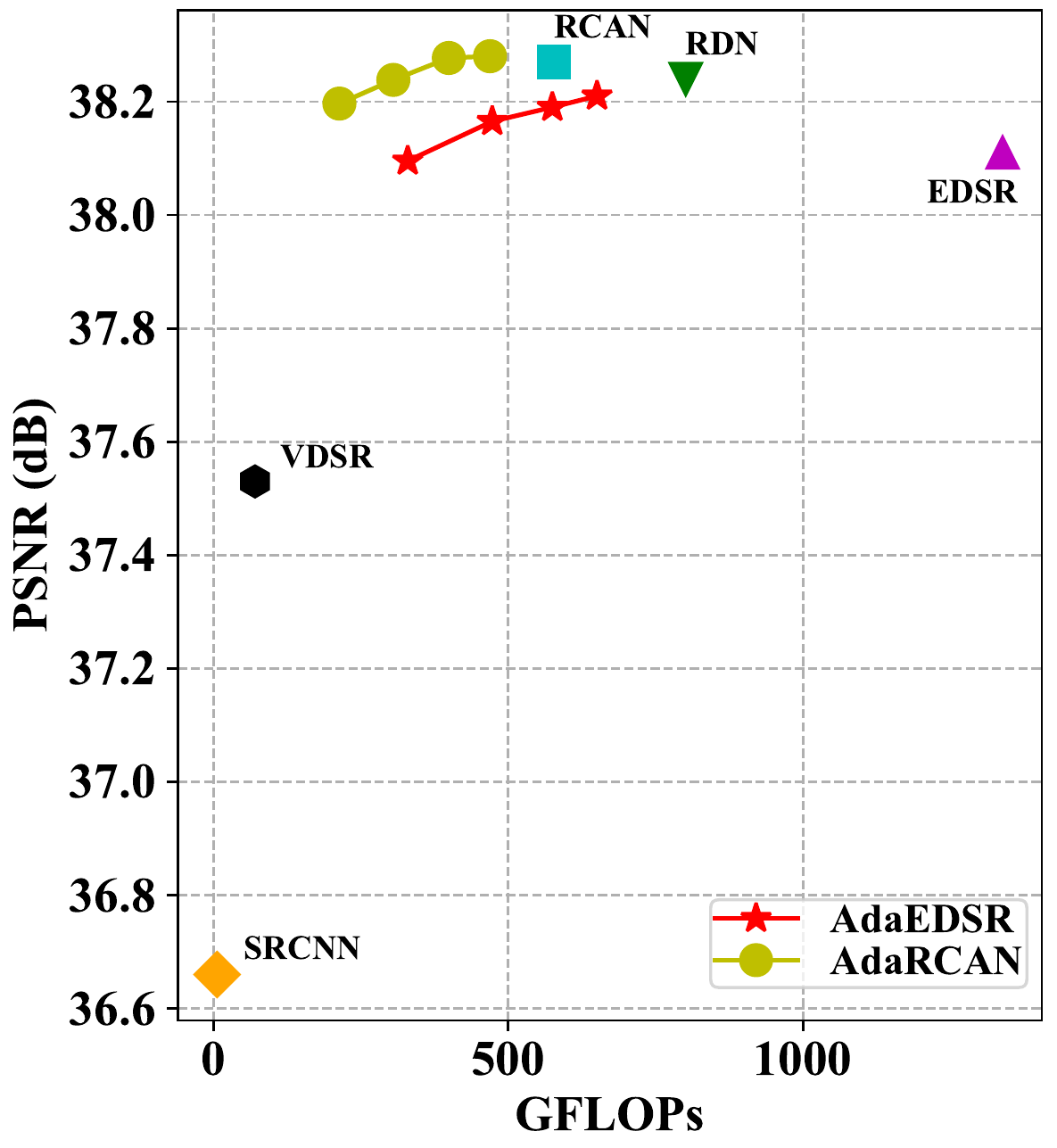}
        \end{minipage}}
    \subfigure[Scale $\times3$]{
    \begin{minipage}{0.3\linewidth}
            \includegraphics[width=.95\linewidth]{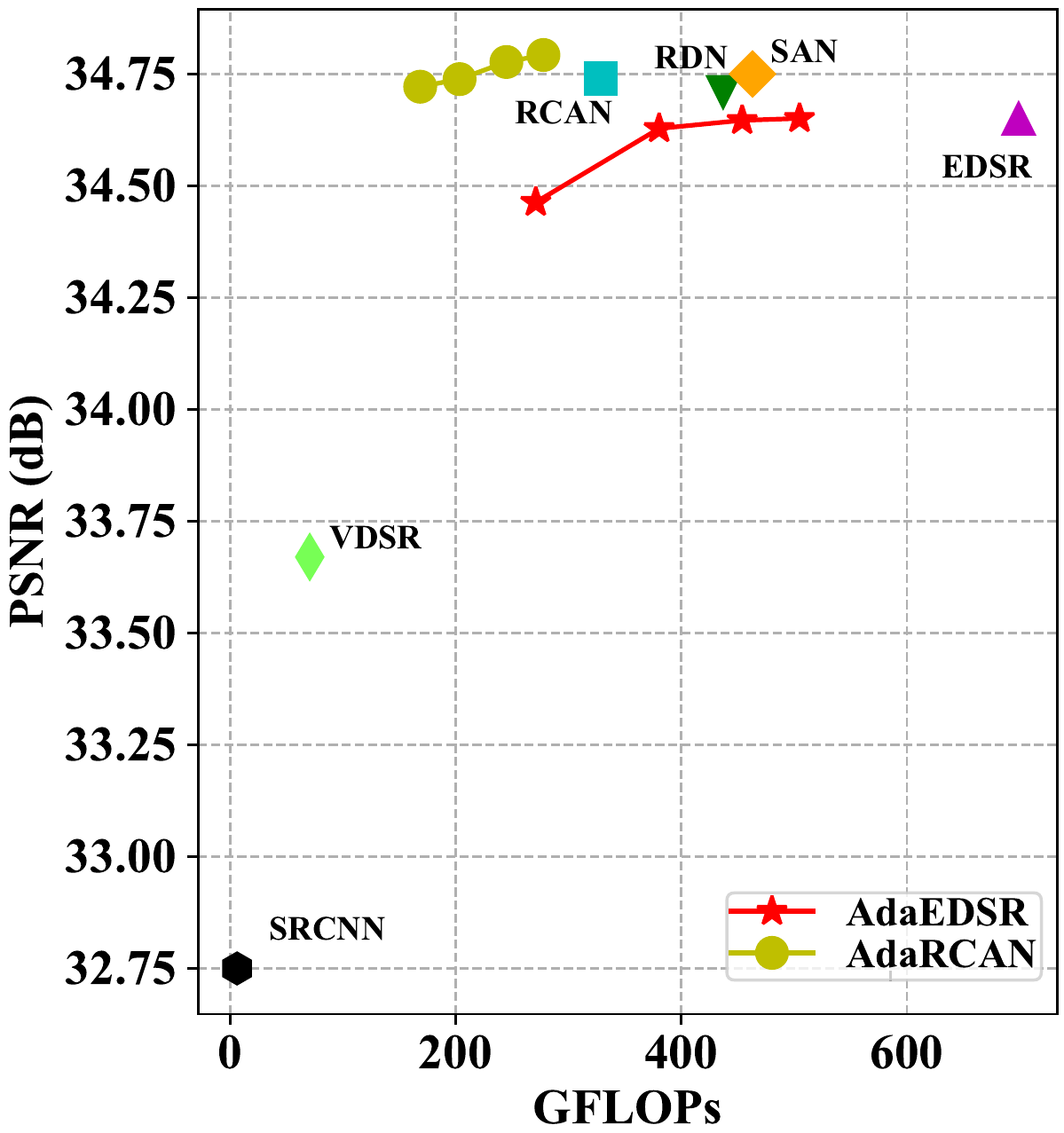}
        \end{minipage}}
    \subfigure[Scale $\times4$]{
    \begin{minipage}{0.3\linewidth}
        \includegraphics[width=.95\linewidth]{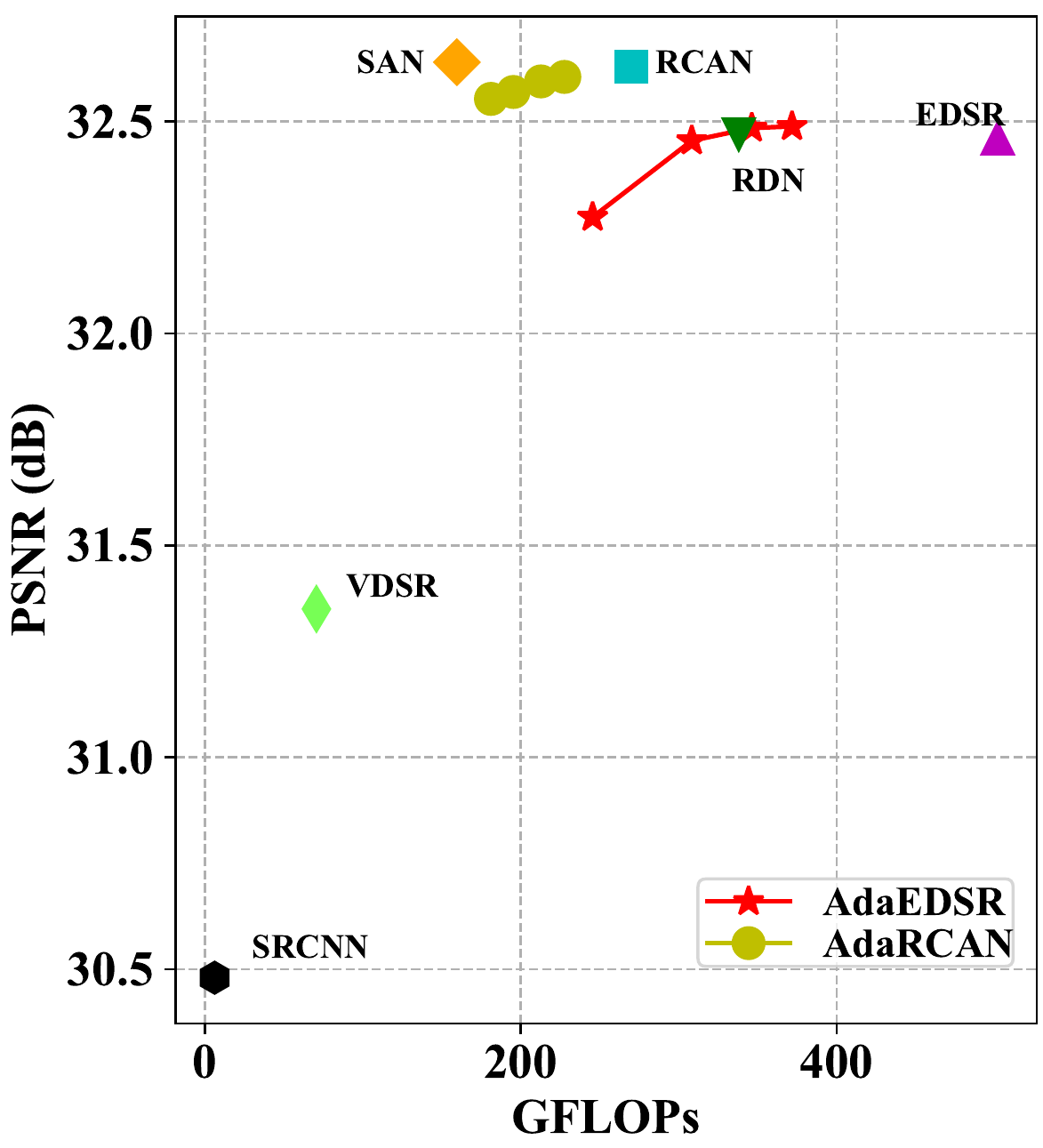}
    \end{minipage}}
    \vspace{-1em}
    \caption{\footnotesize Comparison against state-of-the-art methods in terms of FLOPs and PSNR on Set5. Note that SAN is not given in Scale $\times2$ due to that its computation cost is 3835.9 GFLOPs, which is much more than other methods.}
    \label{fig:adaptive}
    \vspace{-2em}
\end{figure}

Taking both the feature map $\mathbf{z}_0$ and desired depth $d$ as input, the adapter module is able to predict an image content adaptive network depth map while satisfying the computation efficiency constraints.
Consequently, our AdaDSR can be flexibly tuned to meet various efficiency constraints on the fly.
In comparison, the competing methods are based on deterministic inference and can only be performed with the fixed complexity.
As shown in Fig.~\ref{fig:adaptive}, we evaluate our AdaDSR model with different desired depth $d$ (\ie, 8, 16, 24, 32 for AdaEDSR and 5, 10, 15, 20 for AdaRCAN), and record the corresponding FLOPs and PSNR values on Set5. More results please refer to the supplementary materials.

From the figures, we can draw several conclusions.
First, our AdaDSR can be tuned with the hyper-parameter $d$, and resulting in a curve in the figures, rather than a single point as the competing methods.
With an increasing desired depth $d$, AdaDSR requires more computation resources and generates better super-resolved images.
It is worth noting that, AdaDSR taps the potential of the backbone models, and can obtain comparable performance against the well-trained backbone model when higher $d$ is set.
Furthermore, AdaDSR reaches the saturation point with a relatively lower FLOPs, which indicates that a shallower model is sufficient for most regions.
Experiments on both versions (\ie, AdaEDSR and AdaRCAN) verify the effectiveness and generality of our adapter module.

\vspace{-1em}
\section{Ablation Analysis}\label{sec:ablation}
\vspace{-0.7em}
Considering the training efficiency in multi-GPU environment, we perform ablation analysis with EDSR backbone. Without loss of generality, we select AdaEDSR model and scale $\times2$.
\begin{table}
    \centering
    \caption{\footnotesize Quantitative evaluation of EDSR and AdaEDSR variants on Set5 ($\times2$).}
    \label{tab:ablation}
    \scalebox{0.65} {
    \begin{tabular}{|l|c|c|c|c|c|l|c|c|c|c|c|l|c|c|c|c|}
    \cline{1-5}    \cline{7-11}    \cline{13-17}
    \multirow{2}{*}{Method} & PSNR & \multirow{2}{*}{SSIM} & FLOPs & Time & &
    \multirow{2}{*}{Method} & PSNR & \multirow{2}{*}{SSIM} & FLOPs & Time & &
    \multirow{2}{*}{Method} & PSNR & \multirow{2}{*}{SSIM} & FLOPs & Time
    \\
    & (dB) & & (G) & (ms) & &
    & (dB) & & (G) & (ms) & &
    & (dB) & & (G) & (ms)
    \\
    \cline{1-5}    \cline{7-11}    \cline{13-17}

    EDSR (8)
        & 38.05 & 0.9607 & 408.23 & 147.2
    &&
    FAdaEDSR (8)
        & 38.17 & 0.9609 & 504.87 & 280.6
    &&
    AdaEDSR (8)
        & 38.10 & 0.9605 & 329.50 & 169.6
    \\

    EDSR (16)
        & 38.11 & 0.9610 & 718.41 & 230.7
    &&
    FAdaEDSR (16)
        & 38.21 & 0.9611 & 719.62 & 327.7
    &&
    AdaEDSR (16)
        & 38.17 & 0.9608 & 472.90 & 217.0
    \\

    EDSR (24)
        & 38.15 & 0.9612 & 1028.58 & 305.8
    &&
    FAdaEDSR (24)
        & 38.23 & 0.9613 & 997.95 & 366.8
    &&
    AdaEDSR (24)
        & 38.19 & 0.9610 & 574.85 & 243.8
    \\

    EDSR (32)
        & 38.16 & 0.9611 & 1338.76 & 395.7
    &&
    FAdaEDSR (32)
        & 38.24 & 0.9613 & 1358.30 & 402.9
    &&
    AdaEDSR (32)
        & 38.21 & 0.9611 & 650.65 & 312.3
    \\

    \cline{1-5}    \cline{7-11}    \cline{13-17}
    \end{tabular} }
    \vspace{-2em}
\end{table}

\vspace{.3em}
\noindent\textbf{EDSR variants.}
To begin with, we train EDSR variants in our framework, \ie, EDSR~(8), EDSR~(16), EDSR~(24) and EDSR~(32) by setting the number of residual blocks to 8, 16, 24 and 32, respectively.
Note that EDSR~(32) performs slightly better than released EDSR model, so we use this one for a fair comparison.
The quantitative results on Set5 are given in Table~\ref{tab:ablation}.
Comparing all EDSR variants, generally one can observe performance gains as the model depth grows.

Besides, as previously illustrated in Fig.~\ref{fig:intro}(c), a shallow model is sufficient for smooth areas, while regions with rich contexture usually require a deep model for better reconstruction of the details.
Taking advantage of this phenomenon, with lightweight adapter, AdaDSR is able to predict suitable depth for various areas according to difficulty and resource constraints, and achieves better efficiency-performance tradeoff, resulting in the curve at the top left of their corresponding counterparts as shown in Figs \ref{fig:intro}(d) and \ref{fig:adaptive}. Detailed data can be found in Table~\ref{tab:ablation}.

\vspace{.3em}
\noindent\textbf{AdaEDSR variants.}
We further implement several AdaEDSR variants, \ie, FAdaEDSR~(8), FAdaEDSR~(16), FAdaEDSR~(24) and FAdaEDSR~(32), which are trained with a fixed depth $d=$ 8, 16, 24 and 32 respectively, and the adapter module takes only the image features as input.
The models are trained under the same settings (except for the fixed $d$ in the learning objective) with AdaEDSR.
As shown in Table~\ref{tab:ablation}, with the per-pixel depth map, these models obtain much better quantitative results than EDSR variants with similar computation cost.

It is worth noting that FAdaEDSR (32) achieves comparable performance with RDN~\cite{RDN}, which clearly shows the effectiveness of the predicted network depth map.
Furthermore, we also show the performance of our AdaEDSR $\times2$ model in Table~\ref{tab:ablation}. Specifically, AdaEDSR ($n$) means that desired depth $d=n$ at the test time.
One can see that although the quantitative performance is slightly worse than FAdaEDSR, AdaEDSR is more computationally efficient and can be flexibly tuned in the testing phase, indicating that AdaDSR achieves adaptive inference with minor sacrifice of performance.

\vspace{-1.2em}
\section{Conclusion}\label{sec:conclusion}
\vspace{-.7em}
In this paper, we revisit the relationship between the model depth and quantitative performance on single image super-resolution task, and present an AdaDSR model by incorporating a lightweight adapter module and sparse convolution in deep SISR networks.
The adapter module predicts an image content oriented network depth map, and the value is higher in regions with detailed textures and lower in smooth areas.
According to the predicted depth, only a fraction of residual blocks are performed at each point by using im2col based sparse convolution.
Furthermore, the parameters of the adapter module are adjustable on the fly according to the desired depth, so that the AdaDSR model can be tuned to meet various efficiency constraints in the inference phase.
Experimental results show the effectiveness and adaptiveness of our AdaDSR model, and indicate that AdaDSR can obtain state-of-the-art performance while adaptive to a range of efficiency requirements.

\clearpage
%
%
\bibliographystyle{splncs04}
\bibliography{AdaDSR}
\end{document}